\documentclass[11pt,twoside]{article}

\usepackage{asp2004}
\usepackage{epsf}
\usepackage{lscape}

\begin{document}
\title{Light variations and angular momentum loss from the He-strong
magnetic chemically peculiar star HD 37776}   
\author{Z. Mikul\'a\v sek$^{1}$, J.Krti\v{c}ka$^{1}$, J. Zverko$^{2}$,
J. \v{Z}i\v{z}\v{n}ovsk\'y$^{2}$, and J. Jan\'\i k$^{1}$}
\affil{$^1$Institute of Theoretical Physics and Astrophysics,
Masaryk University, Kotla\v{r}sk\'a 2, CZ-611\,37 Brno, Czech
Republic, mikulas@ics.muni.cz}

\affil{$^2$Astronomical Institute of the Slovak Academy of Science,
SK-059\,60 Tatransk\'a Lomnica, Slovak Republic}

\begin{abstract} 
We present a  principle component analysis of multicolour light
variations of the He-strong magnetic chemically peculiar star
HD~37776 and its (O-C) diagram. The period increase rate found,
$\dot{P}/P=5.5\cdot 10^{-6}/\mathrm{year}$, is consistent
with the conception of angular momentum loss through
magnetically confined stellar wind.
\end{abstract}

\section{Introduction}

HD~37776 = HIP~26742 = V901~Ori, B1\,IV, is a member of the well
studied Orion~OB~1 association. The star was recognized as a
helium-strong CP star by Nissen (1976). Pedersen \& Thomsen (1977)
and Pedersen (1979) published 54 excellent \emph{uvby} and 53
photometric measurements of the He\,{\sc i}$\lambda4026$ line of the
star revealing it as both a low amplitude spectrum and light
variable with a rotational period of $1.5385\pm0.003$ days.

Walborn (1982) and Shore \& Brown (1990) reported on the periodic
variability of its H, He, C, Si and perhaps Mg spectral lines.
Thompson \& Landstreet (1985) found an extraordinary double-wave
magnetic curve with a period of 1.53869 days and argued that this
star has a dominantly quadrupolar field geometry. Adelman \& Pyper
(1985) and Adelman (1997) obtained spectrophotometry and two sets of
\emph{uvby} photometry (18 and 42 observations respectively).
Adelman defined the period more precisely and gave the ephemeris:
\begin{eqnarray}
\rm{JD}(\mathit{B}_1^+)=2445724.669(20)+1.538675(5)\cdot \mathit{k},
\end{eqnarray}
$k$ is an integer. The zero phase corresponds to the first magnetic
maximum after Thompson \& Landstreet \cite{thomp}. The star was
photometrically observed in a near infrared region (Catalano \&
Leone 1998, 40 observations in \emph{JHK} colours) and by HIPPARCOS
(103 $H_{\mathrm{p}}$ measurements, ESA 1997).

\section{O-C diagram. New ephemeris}

One of  unsolved problems of the physics of the He~strong/He~weak
mCP stars is the nature of photometric spots on their surfaces which
are the cause of the observed photometric variations of these stars.
We have chosen HD~37776 as a prototype of the He-strong mCP group of
stars as it seems to us to be a relatively simple and at the same
time very well monitored object. A partial step of our research was
to improve the rotational ephemeris of the star, so that it was
applicable to the whole interval of 35 years covered by measurements
of all kind.

For this purpose we used all the satisfactorily accurate photometric
data available, namely the 3 sets of \emph{uvby} observations
(Pedersen \& Thomsen 1977, Adelman \& Pyper 1985 and Adelman 1997)
and $H_{\mathrm p}$ measurements (ESA 1997). This material comprises
559 individual measurements in 5 photometric colours and spans over
20 years. The data were processed by our own code
\mbox{\textbf{PERSYL}} based on the principal component analysis
using the robust regression approaches (Mikul\'a\v{s}ek et al. 2005
and 2003).

\begin{figure}[!ht]
\center \resizebox{0.55\hsize}{!}{\includegraphics{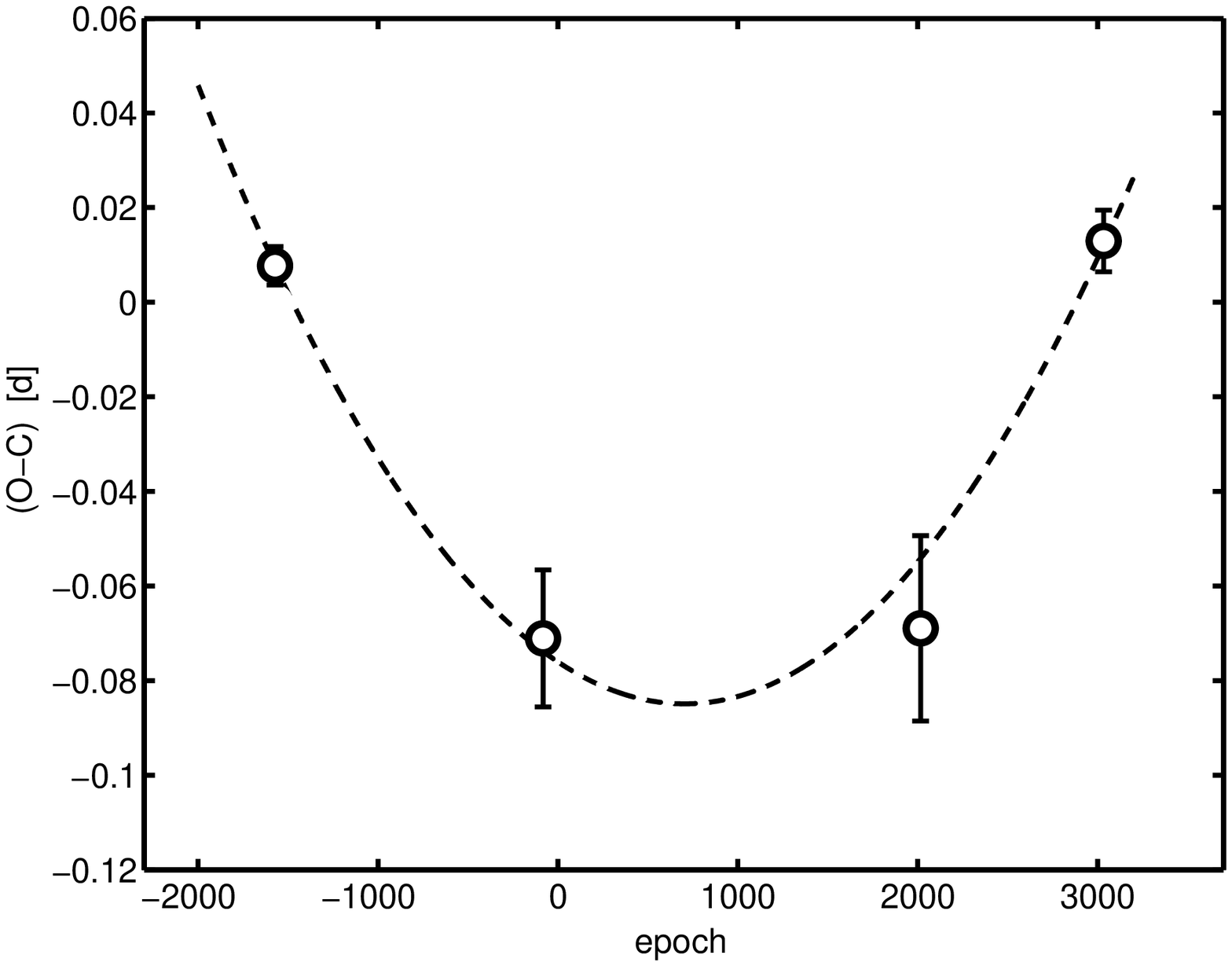}}

\small {\textbf{Fig.\,1}: O-C diagram for HD~37776 versus linear
ephemeris.}
\end{figure}

The departures  we found in the (O-C) {\it vs.} linear light elements
dependence lead us to introducing a quadratic term in to the
ephemeris. Then the moments of the maxima of the light curves,
$\rm{JD_{max}}$, are given by the relation:
\begin{eqnarray}
\mathrm{JD}_{\mathrm{max}}=M_0+P_0\cdot k+\frac{1}{2}\,\dot{P}P_0
\left( k^2-\frac{\overline{E^3}}{\overline{E^2}}\cdot k -
\overline{E^2}\right),
\end{eqnarray}
 where $M_0$ is the moment of the basic maximum
for the $k=0$. $E$ is the epoch, overlines denote the mean weighted
values of a quantity, $P_0$ is the mean period corresponding to
linear approximation, $\dot{P}$ is the derivative of the
instantaneous period, $\overline{E^2}=4\,254\,000$, $
\overline{E^3}/ \overline{E^2}=1409,\ M_0=2\,445\,224.197(19),\
P_0=1.5386762(17)\,\mathrm {days},\
\dot{P}=(23.2\pm3.3)\cdot10^{-9},\
 \dot{P}/P_0=5.5\cdot10^{-6}/\mathrm {year}$. It is very likely the
period is truly increasing as the value of $\dot{P}$ is 7-times
larger than its uncertainty. The mean period $P_0$ is exactly equal
to the Adelman's value but with the error significantly reduced.

Reiners et al. (2000), as well as Oksala \& Townsend (2006) noticed
increase of a period of another magnetic He-strong star: $\sigma$
Ori E. Using the period  given in the latter paper and the value
given in Hesser et al. (1977) we inferred the rate of spindown of
the star: $\dot{P}/P_0=3.2\cdot10^{-6}/\mathrm {year}$, in a good
agreement with the one given by Reiners et al. (2000), i. e.
$2.7\cdot10^{-6}/\mathrm {year}$.

\section{Discussion. Conclusion}

The helium-strong stars represent a relatively small group of early
B stars (B1\,V to B3\,V) that show unusually strong He lines for
their effective temperature. They are believed to represent a
high-temperature extension of the classical magnetic Ap/Bp stars.
The observed chemical peculiarity is probably a consequence of
elemental separation under influence of stellar wind.

Provided that the period change (3) is due to a momentum loss
through a magnetically confined stellar wind, then the period change
is (assuming a rigid-body rotation):
\begin{eqnarray}
{\frac{\dot P}{P_0}=\frac{\xi\, \dot M\,r_\mathrm{cor}^2}{I}} ,
\end{eqnarray}
where $r_\mathrm{cor}$ is the radius of the effective corotation,
$\xi$ is a geometric factor, $I=\eta\,M_*R_*^2$ is the stellar
moment of inertia given by a dimensionless constant $\eta$ and the
stellar mass $M_*$ and radius $R_*$. Let us assume that
$r_\textrm{cor}$ is given by the Alfv\'en radius (Weber \& Davis
\cite{weda}) and that the mass-loss rate corresponds to the value
required by the diffusion theory $\dot{M}\sim 10^{-13}$ M$_\odot
$/year (Michaud et al. \cite{misa}). For HD\,37776 with $\xi=0.1,
\,\,\eta=0.06$ we obtain $r_\mathrm{cor}=10\,000\,R_*$ and the
spin-down rate ${\dot P}/{P}\sim10^{-6}/\mathrm{year}$ consistent
with our finding. For the mass-loss rates anticipated by the
decoupling theory (Hunger \& Groote \cite{hugr}), $\dot{M}\sim
10^{-10}$ M$_\odot /\mathrm {year}$ we obtain the same spin-down
rate since $r_\mathrm{cor}$ is now lower.

The nonzero $\dot{P}$ (if real) however, could also be explained by
e.\,g. the light time effect due to an unresolved secondary. New
photometric and spectroscopic observations of the star that are
expected to help resolving the problem are in progress.

\acknowledgements 
This work was supported by grants VEGA 3014, GA \v{C}R 205/04/1267
and MVTS SR-\v{C}R 128/04

\end{document}